# Comparative studies of plastic scintillator strips with high technical attenuation length for the total-body J-PET scanner

Ł. Kapłon[a,b,c*], J. Baran[a,b,c], N. Chug[a,b,c], A. Coussat[a,b,c], C. Curceanu[d], E. Czerwiński[a,b,c], M. Dadgar[a,b,c], K. Dulski[a,b,c], J. Gajewski[e], A. Gajos[a,b,c], B. Hiesmayr[f], E. Kavya Valsan[a,b,c], K. Klimaszewski[g], G. Korcyl[a,b,c], T. Kozik[a], W. Krzemień[b,c,h], D. Kumar[a,b,c], G. Moskal[b,c,i], S. Niedźwiecki[a,b,c], D. Panek[a,b,c], S. Parzych[a,b,c], E. Pérez del Rio[a,b,c], L. Raczyński[g], A. Ruciński[e], S. Sharma[a,b,c], S. Shivani[a,b,c], R. Shopa[g], M. Silarski[a,b,c], M. Skurzok[a,b,c], E. Stępień[a,b,c], F. Tayefi Ardebili[a,b,c], K. Tayefi Ardebili[a,b,c], W. Wiślicki[g], P. Moskal[a,b,c]

[a] *Faculty of Physics, Astronomy and Applied Computer Science, Jagiellonian University, Lojasiewicza 11, 30-348 Krakow, Poland*
[b] *Total-Body Jagiellonian-PET Laboratory, Jagiellonian University, Lojasiewicza 11, 30-348 Krakow, Poland*
[c] *Center for Theranostics, Jagiellonian University, Krakow, Poland*
[d] *INFN, Laboratori Nazionali di Frascati, 00044 Frascati, Italy*
[e] *Institute of Nuclear Physics, Polish Academy of Sciences, Radzikowskiego 152, 31-342 Krakow, Poland*
[f] *Faculty of Physics, University of Vienna, 1090 Vienna, Austria*
[g] *Department of Complex Systems, National Centre for Nuclear Research, 05-400 Otwock-Swierk, Poland*
[h] *High Energy Physics Division, National Centre for Nuclear Research, 05-400 Otwock-Swierk, Poland*
[i] *Faculty of Chemistry of the Jagiellonian University, Gronostajowa 2, 30-387 Krakow, Poland*

*Corresponding author.
E-mail address: lukasz.kaplon@uj.edu.pl (Łukasz Kapłon).

ABSTRACT

Plastic scintillator strips are considered as one of the promising solutions for the cost-effective construction of total-body positron emission tomography (PET) system. The purpose of the performed measurements is to compare the transparency of long plastic scintillators with dimensions 6 mm × 24 mm × 1000 mm and with all surfaces polished. Six different types of commercial, general purpose, blue-emitting plastic scintillators with low attenuation of visible light were tested, namely: polyvinyl toluene-based BC-408, EJ-200, RP-408, and polystyrene-based Epic, SP32 and UPS-923A. For determination of the best type of plastic scintillator for total-body Jagiellonian positron emission tomograph (TB-J-PET) construction, emission and transmission spectra, and technical attenuation length (TAL) of blue light-emitting by the scintillators were measured and compared. The TAL values were determined with the use of UV lamp as excitation source, and photodiode as light detector. Emission spectra of investigated scintillators have maxima in the range from 420 nm to 429 nm. The BC-408 and EJ-200 have the highest transmittance values of about 90% at the maximum emission wavelength measured through a 6 mm thick scintillator strip and the highest technical attenuation length reaching about 2000 mm, allowing assembly of long detection modules for time-of-flight (TOF) J-PET scanners. Influence of the 6 mm x 6 mm, 12 mm x 6 mm, 24 mm x 6 mm cross-sections of the 1000 mm long EJ-200 plastic scintillator on the TAL and signal intensity was measured. The highest TAL value was determined for samples with 24 mm x 6 mm cross-section.

## 1. Introduction

Advent of total-body positron emission tomography (TB-PET) systems [1], [2], [3], [4] opens new perspectives for medical diagnosis enabling simultaneous imaging of all organs and tissues of the patient [5], [6]. However, the high costs of total-body PET systems based on crystal detectors limit their dissemination in the hospitals and research facilities [4], [6], [7], [8]. One of the possible solution for the cost-effective construction of the TB-PET is proposed by the J-PET Collaboration which is developing TB-PET system based on plastics scintillators [4], [5], [9]. Plastic scintillators manufactured by bulk polymerization method [10], [11] are used as a part of radiation detectors in TB-J-PET scanner [9], [12], [13], [14], [15], [16], [17]. The multiphoton J-PET scanners constructed from plastic scintillators are capable of imaging positronium for medical purposes [4], [18], [19] and detecting ortho-positronium decays for physics experiments [20], [21], [22]. Fast timing properties [23] and high transparency [24] of the plastic scintillators enable building modular [12] TB-J-PET scanners [13] with long axial field-of-view [14]. Polymer-based scintillators with blue [25] emission can be also used as a light signal generating optical materials in plastic scintillation dosimetry [26].

Light emitted in long aspect ratio scintillators is reflected from its surfaces along the scintillator strip via total internal reflection mechanism in similar way as light propagates in optical fibers. Technical attenuation length (TAL) describes signal attenuation of light emitted in the scintillator [24]. Light signal after passing its TAL value along the scintillator is attenuated to 1/e of its initial intensity. The mathematical constant "e" in 1/e is the base of the natural logarithm approximately equals to 2.718. The TAL value depends on transparency of the scintillator for the light which it is emitting, quality of scintillator surface polishing, thickness, and shape of the scintillator and presence of light reflectors attached to the surface of the scintillator. TAL determines maximum useful length of the plastic scintillators that can be applied for construction of radiation detectors, for example in the J-PET scanner [6]. High TAL value increases the amount of optical photons reaching light detectors mounted at both ends of the plastic scintillator and as a consequence increases the time resolution of the total-body J-PET scanner [4]. The TAL values of two most widely used plastic scintillators can be found in Table 1. BC-408 and EJ-200 are manufactured by two different companies but possess the same properties according to data sheets.

Table 1. TAL values for BC-408 and EJ-200 plastic scintillators with rectangular cross-sections and different dimensions (length x width x thickness) found in the literature. High TAL values allows to build long units of time-of-flight detectors. The average TAL value is equal to 216 cm for the BC-408 and 207 cm for the EJ-200.

| Scintillator dimensions (mm) | Surface condition | Excitation source | TAL (cm) | Ref. |
|---|---|---|---|---|
| **BC-408** | | | | |
| 3000 x 40 x 40 | polished | 2 GeV/c kaon beam | 170 | [27] |
| 2790 x 40 x 40 | wrapped with white paper | 2 GeV/c kaon and pion beam | 330 | [28] |
| 2020 x 72 x 72 | wrapped with ESR | cosmic rays | 260 | [29] |
| 2000 x 85 x 50 | diamond milled | 1 GeV/c kaon beam | 280 | [30] |
| 2000 x 40 x 5 | polished, wrapped with Teflon and Tedlar | 2 GeV/c proton, pion and positron beam | 141 | [31] |
| 2000 x 40 x 10 | | | 179 | |
| 2000 x 40 x 20 | | | 207 | |
| 2000 x 40 x 40 | | | 313 | |
| 150 x 50 x 2 | wrapped with aluminium foil | 2 GeV/c pion beam | 66 | [32] |
| 250 x 150 x 2 | | | 66 | |
| 1200 x 150 x 40 | | | 400 | |
| 200 x 20 x 1 | polished | 1.5 GeV/c pion beam | 30 | [33] |
| **EJ-200** | | | | |
| 3000 x 110 x 25 | polished | 10 GeV/c muon beam | 184 | [34] |
| 2300 x 60 x 50 | polished and wrapped with different foils | 0.8 GeV/c electron beam | 246-321 | [35] |
| 2000 x 25 x 25 | polished and wrapped with different foils | 40 GeV/c pion beam | 134-143 | [36] |
| 2000 x 50 x 50 | | | 217-292 | |
| 2000 x 60 x 25 | polished and diamond cut | 5 GeV/c pion beam | 164-220 | [37] |
| 1800 x 80 x 30 | wrapped with aluminized mylar | cosmic rays, $^{90}$Sr | 212 | [38] |
| 900 x 34 x 30 | wrapped with ESR VM-2000 and Tedlar | 10 GeV/c electron beam | 140 | [39] |

Emission spectra of scintillators should overlap with maximum detection efficiency of used light detectors [23], [26]. Majority of commercial plastic scintillators have maximum of emission spectra positioned from 370 nm to 580 nm (data from catalogues of companies listed in subsection 2.1) which is in the range of maximum detection efficiency of standard photomultiplier tubes. Overlap of emission spectrum of plastic scintillator with absorption spectrum of wavelength shifter bars is preferred in radiation detectors for the precise determination of interaction points in scintillating material [40]. Shapes and maxima of emission spectra of BC-408 and EJ-200 scintillators do not change after damaging by gamma rays [41]. However, shape, intensity and maximum of wavelength spectrum change along the scintillator fibers [42]. Short-wavelength part of spectrum is attenuated, spectrum intensity is decreasing and maximum of the emission spectrum is shifting towards longer wavelengths as distance between excitation point and photodetector is increasing [43].

Transmission spectra of plastic scintillators are measured for example to determine degree of radiation damage after gamma rays irradiation [41]. Transparency of scintillating material depends mainly on scintillator type, thickness, wavelength and surface polishing. Transmittance at the wavelength of maximum emission depends on scintillator type and thickness: for BC-408 and EJ-200 scintillators transmittance is about 70% for 60 mm thick samples [41], EJ-200 scintillator possess transmittance of 85% for few millimeter thick sample [44], BC-408 and UPS-923A scintillators have up to 90% transmittance for 4 mm thick samples [45], UPS-923A scintillator has 84% transmittance for 6 mm thick sample [46], RP-408 scintillator has 89% transmittance for 10 mm thick sample [26].

The purpose of this research is to measure the optical properties of selected plastic scintillators commonly used for assembling TOF radiation counters. Many plastic scintillators possess very similar scintillation properties, see Table 2, but TAL value is not given by the manufacturers. Instead producers provide so called bulk attenuation length (BAL), which depends on the light attenuation in bulk volume of the scintillator only [47]. BAL describes the maximum transparency of scintillators, and its value is higher than the value of TAL, therefore BAL is more often provided by manufacturers in data sheets. BAL does not includes results of any effects that decrease amount of light intensity reaching the end of scintillator, such as losses due to light reflections from scintillators surfaces and impact of cross-section and thickness of scintillators on the light attenuation. TOF radiation detectors usually use long plastic scintillators where the light propagates inside of the scintillator with many total internal reflections. Therefore, determining TAL is more useful than BAL, and additionally, the TAL can be confirmed by measurement of scintillators transmission spectra through selected thickness of the scintillator. The TAL value can be correlated with the transmission spectra because transparency of bulk scintillator material is important factor influencing the TAL. Emission spectra, transmission spectra through 6 mm thick strip, and TAL values of 1000 mm long six different types of plastic scintillators were measured.

## 2. Materials and methods

### 2.1. Scintillator samples

Six types of plastic scintillators produced by different manufacturers were tested:
- BC-408 from Saint-Gobain Crystals (USA),
- EJ-200 from Eljen Technology (USA),
- RP-408 from Rexon Components (USA),
- the plastic scintillator from Epic-Crystal (China),
- SP32 from Nuviatech Instruments (Czechia),
- UPS-923A from Amcrys, Institute for Scintillation Materials (Ukraine).

BC-408, EJ-200 and RP-408 are scintillators with the same scintillation properties according to data sheets, and are made from polyvinyl toluene (PVT) as polymer base. Epic, SP32 and UPS-923A are manufactured from polystyrene (PS), and possess similar properties. The mentioned scintillators were chosen for the investigation because they combine high light output, short decay time and high light attenuation length, see Table 2. Additionally, wavelengths of maximum emission for the researched scintillators are centered in the range from 418 nm to 425 nm. These maxima of emission spectra are close to maximum of quantum efficiency of standard photomultiplier tubes (PMT) with bialkali photocathode located at 420 nm and peak sensitivity wavelength of silicon photomultipliers (SiMP) centered at 450 nm. Both types of light detectors mentioned before are widely used in fast TOF radiation detectors and PET scanners.

Table 2. Properties of the plastic scintillators based on data sheets. The light attenuation length is given as BAL.

| Plastic scintillator | Light output (ph/MeV) | Decay time (ns) | Wavelength of maximum emission (nm) | Bulk light attenuation length (cm) | Polymer base | Refractive index | Density (g/cm$^3$) |
|---|---|---|---|---|---|---|---|
| BC-408 | 10 000 | 2.1 | 425 | 380 | PVT | 1.58 | 1.023 |
| EJ-200 | 10 000 | 2.1 | 425 | 380 | PVT | 1.58 | 1.023 |
| RP-408 | 10 000 | 2.1 | 425 | 400 | PVT | 1.58 | 1.032 |
| Epic | 8 600 | 2.4 | 415 | 200 | PS | 1.58 | 1.050 |
| SP32 | 8 750 | 2.5 | 425 | - | PS | 1.57 | 1.030 |
| UPS-923A | 8 750 | 3.3 | 418 | 400 | PS | 1.60 | 1.060 |

To determine the TAL value of the six scintillator materials, we have tested bars with rectangular cross-section and dimensions of 6 mm x 24 mm x 1000 mm with all surfaces polished by manufacturers. The 6 mm × 24 mm cross-section of the scintillators was chosen to match the surface of 1x4 SiPM array with an active surface of 6 mm x 6 mm per one photomultiplier attached at both ends of the scintillator [15]. The assumed length of the scintillator (1000 mm) corresponds to the maximal dimension of one module in total-body J-PET scanner. For the TAL measurements of scintillators with different dimensions, three rectangular cross-sections were investigated: 6 mm x 6 mm, 12 mm x 6 mm, 24 mm x 6 mm, and length of 1000 mm for each cross-sections. The plastic scintillators with smaller cross-sections were considered in this study because segmenting of the scintillator into smaller pieces can increase spatial resolution of the J-PET scanner.

Quality of surface polishing is different in all tested scintillators. BC-408 and EJ-200 have two large surfaces (so called faces, 24 mm x 1000 mm) as-cast and all other edges diamond-milled. UPS-923A has all surfaces milled and polished by buffing wheel. The rest of scintillators have all surfaces polished by the buffing wheel or similar technique. Investigated scintillators were used as received from manufacturers, the scintillators have original surfaces polished by the producers, and no reflective foils were used for the measurements. The plastic scintillators were not additionally polished in our laboratory since this approach would be impractical in case of ordering larger quantities due to the fact that polishing is a time consuming process, and its quality cannot be as good as done in the factory on special equipment.

### 2.2. Measurements

Emission spectra of the studied scintillators were measured using the USB4000 fiber optic spectrometer produced by Ocean Optics. Scintillator strips were excited by ultraviolet (UV) lamp with maximum emission centered at 365 nm. Ocean Optics P400-2-UV-VIS quartz-core optical fiber transmitting emission signals from the scintillators surface to the spectrometer was positioned at a right angle (an angle of 90 degree) to the scintillator surface. A right angle between scintillator surface with optical fiber and scintillator surface excited by UV lamp light was used, see scheme in figure 1. Distance from fiber and UV spot was few mm. To determine wavelength of maximum emission, part of the Gaussian function was fit to the emission peak of the spectrum and its maximum was obtained. The full width at half maximum (FWHM) and full width at tenth maximum (FWTM) of the emission spectra were calculated by subtracting the wavelength values from the left and right sides of the normalized spectrum corresponding to the half and the tenth of its maximum, respectively.

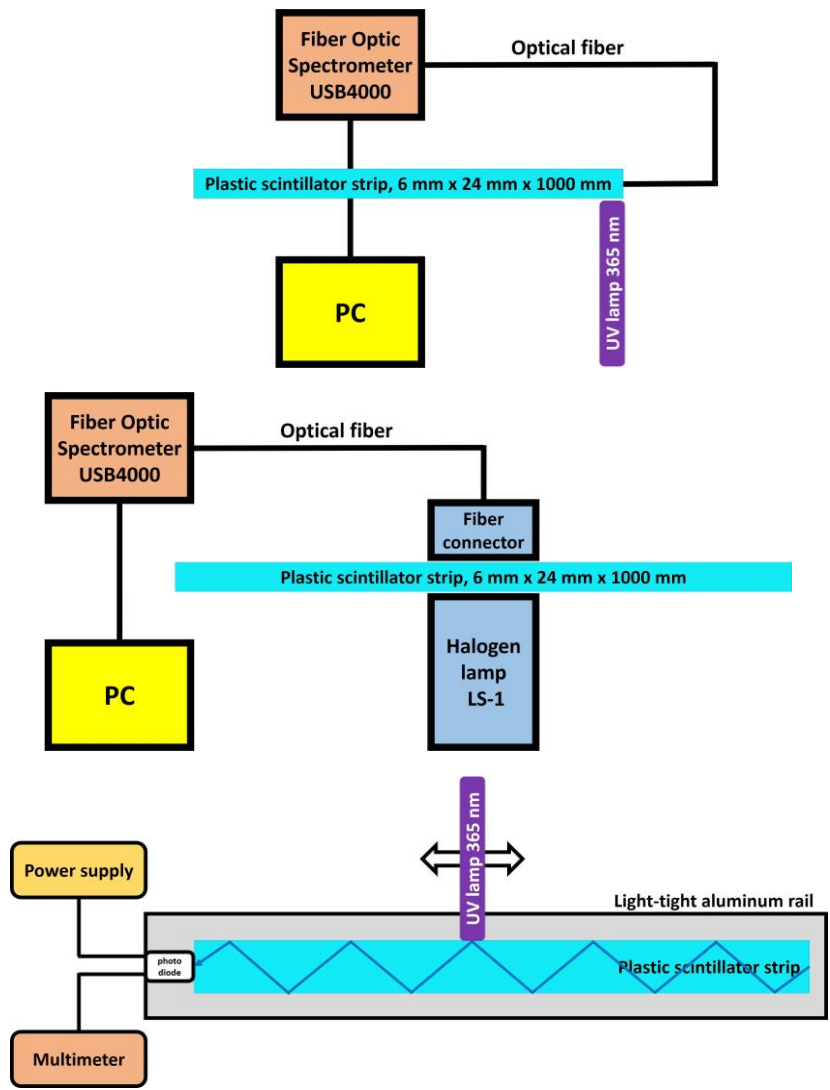

Figure 1. Schemes of experimental setups used for measuring of emission spectra (top panel), transmission spectra (middle panel) and technical attenuation length and average voltage signal (bottom panel) of plastic scintillators.

Transmission spectra, relative to the air, were measured with the same USB4000 spectrometer and using the same optical fiber. LS-1 tungsten halogen lamp from Ocean Optics was used as a broadband light source. Scintillators with 6 mm thickness were horizontally inserted into a slit (6.4 mm wide and 40 mm long) in fiber connector in the LS-1 lamp, see Figure 1. To minimize influence of the small surface imperfections on the transmission spectra, for each scintillator strip five measurements of the transmission spectra spaced about 200 mm apart along the length of the scintillator were measured. The final transmission spectrum was averaged over the five measurements performed for each scintillator type.

For light attenuation length measurements scintillators were excited by 5 mm diameter spot of UV light emitted from lamp with 365 nm emission maximum. The diameter of the UV spot (5 mm) was smaller than the minimum width of the tested scintillator (6 mm) to avoid misalignment errors during shifting the UV spot along the surface of the scintillator. The UV spot was always positioned at the center of 24 mm x 1000 mm large surface of the scintillator along the surface axis. Part of blue light emitted in the scintillator travels to its both ends by total internal reflections. Photodiode with enhanced blue sensitivity Osram BPW 34B, air-coupled at one end of the scintillator, converts the blue light into electrical signals which were measured by a digital multimeter Rigol DM3064, Rigol Technologies. The photodiode has increasing spectral sensitivity in range from 350 nm to its maximum value at 850 nm, than sensitivity is decreasing in range from 850 nm to 1100 nm. In the investigated region with emission of scintillators centered from 420 nm to 430 nm, increase of this sensitivity is small. Scintillators have emission spectra with similar shape, range and maxima located around 425 nm, and that sensitivity should not influence obtained results. Scintillator-photodiode air coupling was used because measurements were easier and faster. Optical gel was also avoided because it forms hard to remove layer on the scintillator surface which was intended for gluing in the future experiment. The plastic scintillator was positioned in a light-tight aluminum rail. Measurement was performed in 50 mm steps. Spot of the UV light excites scintillator through holes in the aluminum rail.

The TAL value was obtained by fitting (in Origin software) sum of two exponential functions to the data points, as described in our previous research [24]:

$$I(x) = A_1 \times e^{\left(-\frac{x}{\lambda_S}\right)} + A_2 \times e^{\left(-\frac{x}{\lambda_L}\right)} + y_0 \tag{1}$$

where $\lambda_S$ and $\lambda_L$ are the short and long attenuation length components, $A_1$ and $A_2$ are amplitudes, and $y_0$ denotes constant background from photodiode and multimeter noise without light in the setup. The TAL value is represented by the long attenuation length component, $\lambda_L$. The short attenuation length component, $\lambda_S$, represents highly attenuated part of emission spectrum and direct light incoming to the photodiode without total internal reflections. The short component of the TAL dominates in short distance in the first few centimeters between excitation UV lamp and photodiode, where blue light from the scintillator decrease rapidly. After about 20 cm of the UV lamp-photodiode distance, the second TAL component dominates, and decrease of the blue light emitted by the scintillator is less steep. In Figures 4 and 5, experimentally obtained the shape and steepness of the decrease of the light signal along 1000 mm long plastic scintillators are shown.

For each scintillator type, four measurements of TAL were performed on two pieces of the same scintillator. Scanning by moving UV lamp into holes in light-tight rail in 50 mm steps of scintillator strip along the whole length and in the reversed direction were done for two pieces of the same scintillator type. The four TAL results for each scintillator type were averaged as shown in Table 3. Standard deviation from the four measurements was given as a statistic uncertainty.

For the cross-section investigation, five EJ-200 plastic scintillator strips of the following dimensions 6 mm x 6 mm, 12 mm x 6 mm, 24 mm x 6 mm were tested. Each strip was measured twice: from one end to the other and in reverse direction. From the ten measurements for each cross-sections, average and standard deviation of TAL were calculated. The tests were performed by measuring amplitude of the voltage signals for each position of the UV lamp used to excite the scintillation light. The UV lamp was shifted along the scintillator in steps of 5 cm. Voltage signal amplitudes for 5 cm photodiode-excitation point distance were extracted from two measurements of each five EJ-200 scintillators for each cross-section, data were averaged and standard deviation was calculated.

## 3. Results and discussion

### 3.1. Emission spectra

The emission spectra of the investigated plastic scintillators are shown in Figure 2. Wavelengths of maximum emission and full width at half maximum (FWHM) of emission spectra are listed in Table 3.

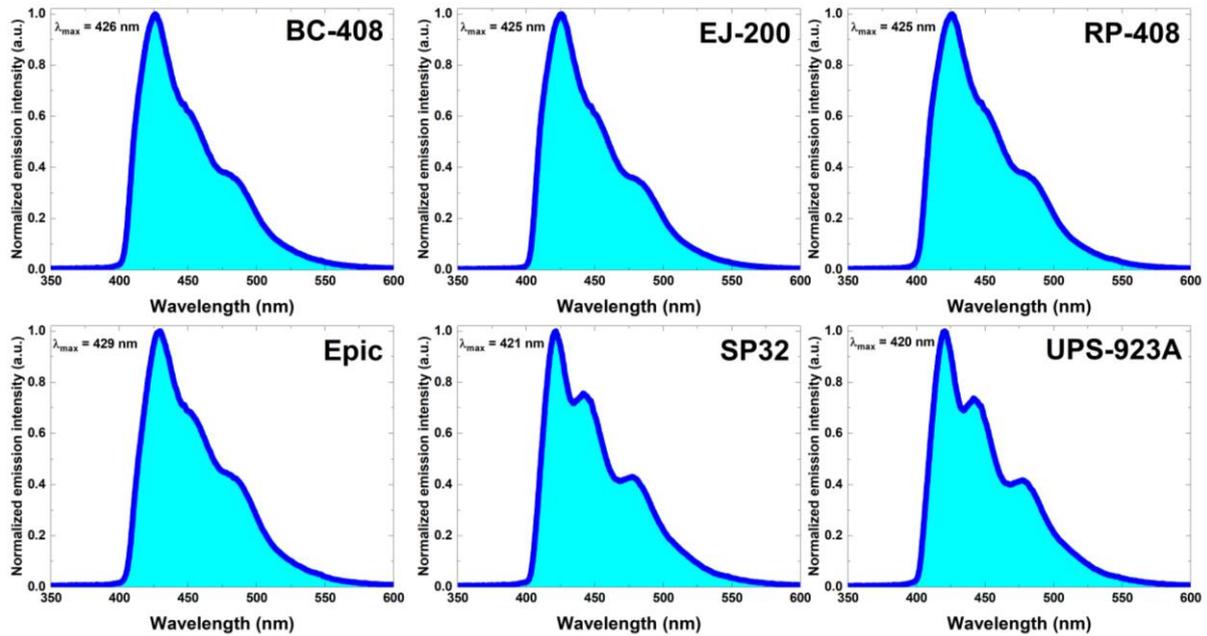

Figure 2. Emission spectra of the investigated plastic scintillators measured in this work. All spectra have emission wavelength maxima positioned in range from 420 nm to 429 nm (blue light).

Wavelength of maximum emission for PVT-based scintillators equals 425 - 426 nm, and shape of the spectra are the same. This result and the same scintillation properties of these three scintillators listed in Table 2, suggest that BC-408, EJ-200 and RP-408 contain the same wavelength shifter (WLS), i.e. fluorescent substance emitting blue light in the scintillation composition. PS-based SP32 and UPS-923A scintillators have the same wavelength of maximum emission which equal to 420 - 421 nm, and shape of the spectra with additional distinct peaks around 442 nm and 478 nm. Taking into account that UPS-923A scintillator contains 1,4-bis(5-phenyl-2-oxazolyl)benzene, POPOP fluorescent dye, as WLS [47], it can be assumed that also SP32 contains the same POPOP WLS.

Table 3. Properties of the plastic scintillators with dimensions 6 mm x 24 mm x 1000 mm measured in this work. Transmittance at the wavelength of maximum emission was measured through 6 mm thick scintillator strip. TAL was measured four times for each scintillator, and uncertainties represent the standard deviation of the four measurements.

| Plastic scintillator | Wavelength of maximum emission (nm) | FWHM of emission spectrum (nm) | FWTM of emission spectrum (nm) | Transmittance at the wavelength of maximum emission (%) | Technical attenuation length (cm) |
|---|---|---|---|---|---|
| BC-408 | 426.08 ± 0.02 | 51.3 | 114.5 | 90.5 ± 0.2 | 205.7 ± 26 |
| EJ-200 | 425.32 ± 0.03 | 50.5 | 113.1 | 90.1 ± 0.2 | 190.9 ± 4.9 |
| RP-408 | 425.45 ± 0.03 | 52.8 | 116.0 | 66.2 ± 4.3 | 52.6 ± 4.5 |
| Epic | 428.99 ± 0.03 | 54.6 | 119.0 | 76.1 ± 1.7 | 45.9 ± 4.4 |
| SP32 | 421.35 ± 0.03 | 47.5 | 118.2 | 76.8 ± 0.5 | 28.8 ± 0.4 |
| UPS-923A | 420.84 ± 0.02 | 48.2 | 119.1 | 83.9 ± 0.3 | 73.4 ± 8.4 |

The highest wavelength of maximum emission equals 429 nm for the Epic scintillator differs significantly from the emission maximum of 415 nm quoted in a data sheet. This difference could result from lower transmittance of light for the Epic scintillator, see subsection 3.2. In the case where scintillation material has lower transmittance at the wavelength of maximum emission, short-wavelength part of the emission spectrum is attenuated within few millimeters and the wavelength of maximum emission is shifted to the longer values. FWHM and FWTM of emission spectra for all the tested scintillators amount to about 51 nm and 117 nm, respectively, but their shape is significantly different. SP32 and UPS-923A have two additional local maxima at 442 nm and 478 nm while the rest of scintillators have more smooth shape.

For all the tested materials, except Epic scintillator, measured emission maxima are in good agreement with the data sheets values. Few nanometers differences are caused by different spectrometers and excitation sources used, and its geometry, signal readout and scintillator shape. Shift of emission spectrum into longer wavelength for Epic scintillator can be caused by lower optical transmittance of base polymer (Table 3) which strongly attenuates shorter wavelength of the emission spectrum. Overlapping of absorption and emission spectra of blue-emitting fluorescent substance (self-absorption) can shift emission maximum into longer wavelength. In polystyrene scintillators absorption coefficient increases for shorter wavelengths, especially for wavelengths shorter than maximum of emission spectrum [48].

### 3.2. Transmission spectra

The measured transmission spectra are presented in Figure 3. Manufacturers of plastic scintillators do not quote this information in data sheets. Sharp rise of the transmittance at about 400 nm is connected to emission spectra which also start near 400 nm. Below the left edge of the emission spectrum light is absorbed by the used WLS in chemical composition of the scintillator. The transmission spectra flatten above the wavelength of maximum emission where WLS fluorescent dye emits light and does not absorb it.

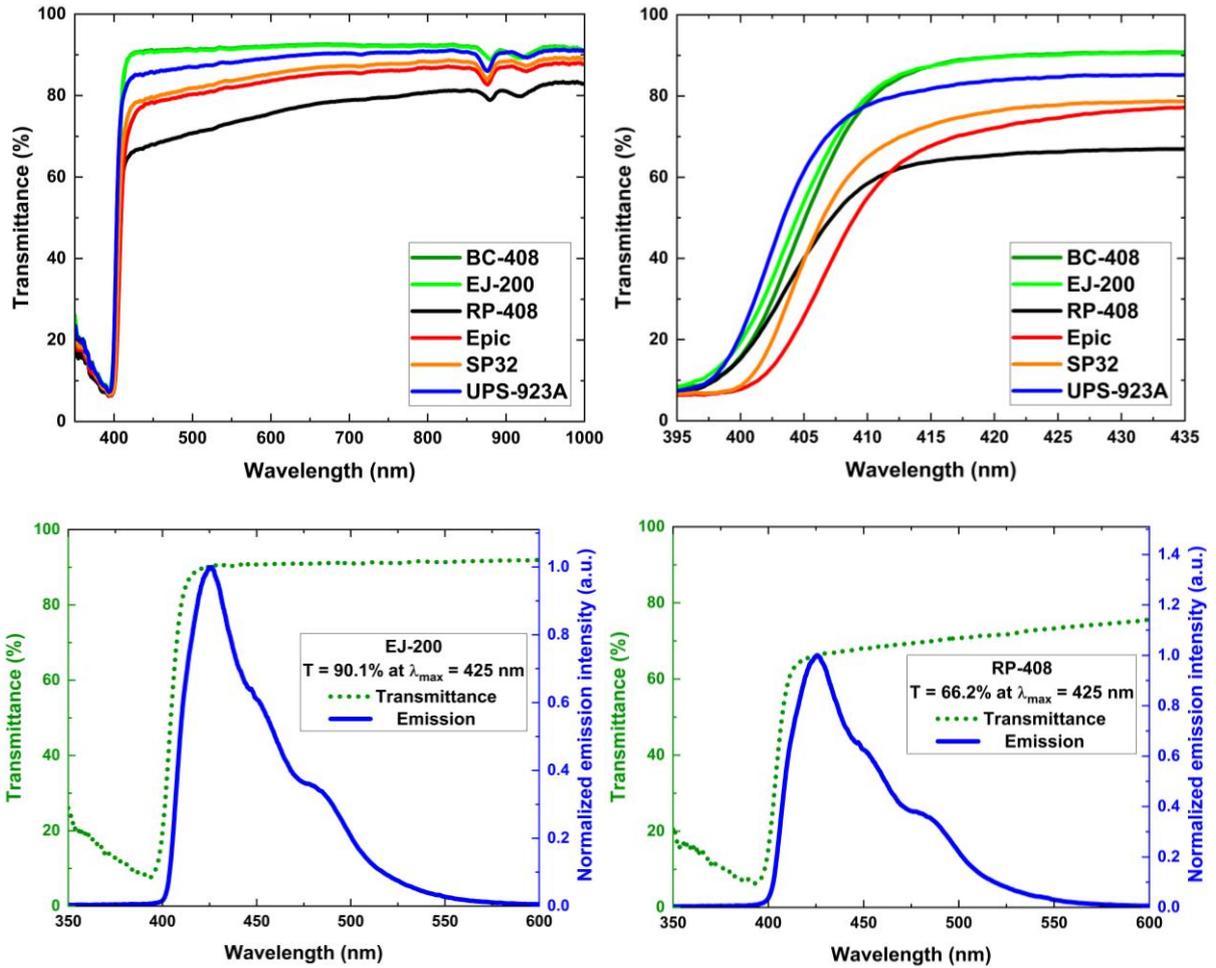

Figure 3. Transmission spectra of the plastic scintillators with 6 mm thickness measured in this work: in full range of the spectrometer (top left panel) and the same spectra with magnified wavelength range where maxima of the emission spectra are positioned (top right panel). Each of the spectra for the six types of the scintillators were averaged with five measurements performed along a 1000 mm long strip every 200 mm. Transmission spectra of one of the most transparent EJ-200 and one of the least transparent RP-408 plastic scintillators superimposed on its emission spectra (bottom panels) to indicate transmittance at the wavelength of maximum emission point.

Transmittance of the scintillating material is a sum of transparency of the material through the volume of scintillator (this depends on the polymer transparency, the type and concentration of fluorescent additives, self-absorption of fluorescent dyes, and the presence of impurities absorbing UV and visible light [49]), thickness of the scintillator, reflections at both surfaces of the scintillator, and scattering of the light on optical defects (scratches, pores, semi-circular patterns from milling) in the surfaces and inside of the scintillators volume. The best transmittance through 6 mm thick plastic scintillators we have determined for BC-408 and EJ-200 samples obtaining value of about 90%. This high transmittance value results from the fact that BC-408 and EJ-200 have large surfaces (24 mm x 1000 mm) as-cast, with high optical quality for light transmission. The UPS-923A has slightly smaller value of about 84% transmittance, which caused by the machining process and other type of polishing than as-cast scintillators. The smallest transmittance of 66% possess RP-408 sample which has surface full of scratches from buffing wheel polishing method.

### 3.3. Technical attenuation length

Results of the TAL measurements for six types of commercially available plastic scintillators strips investigated in this work are presented in Figure 4. In the top-right and bottom-right panels of Figure 4, an example of fitting sum of two exponential functions to the data points for BC-408 scintillator with the highest transparency to the emitted light, and for SP32 scintillator with the lowest transparency to the emitted light are presented.

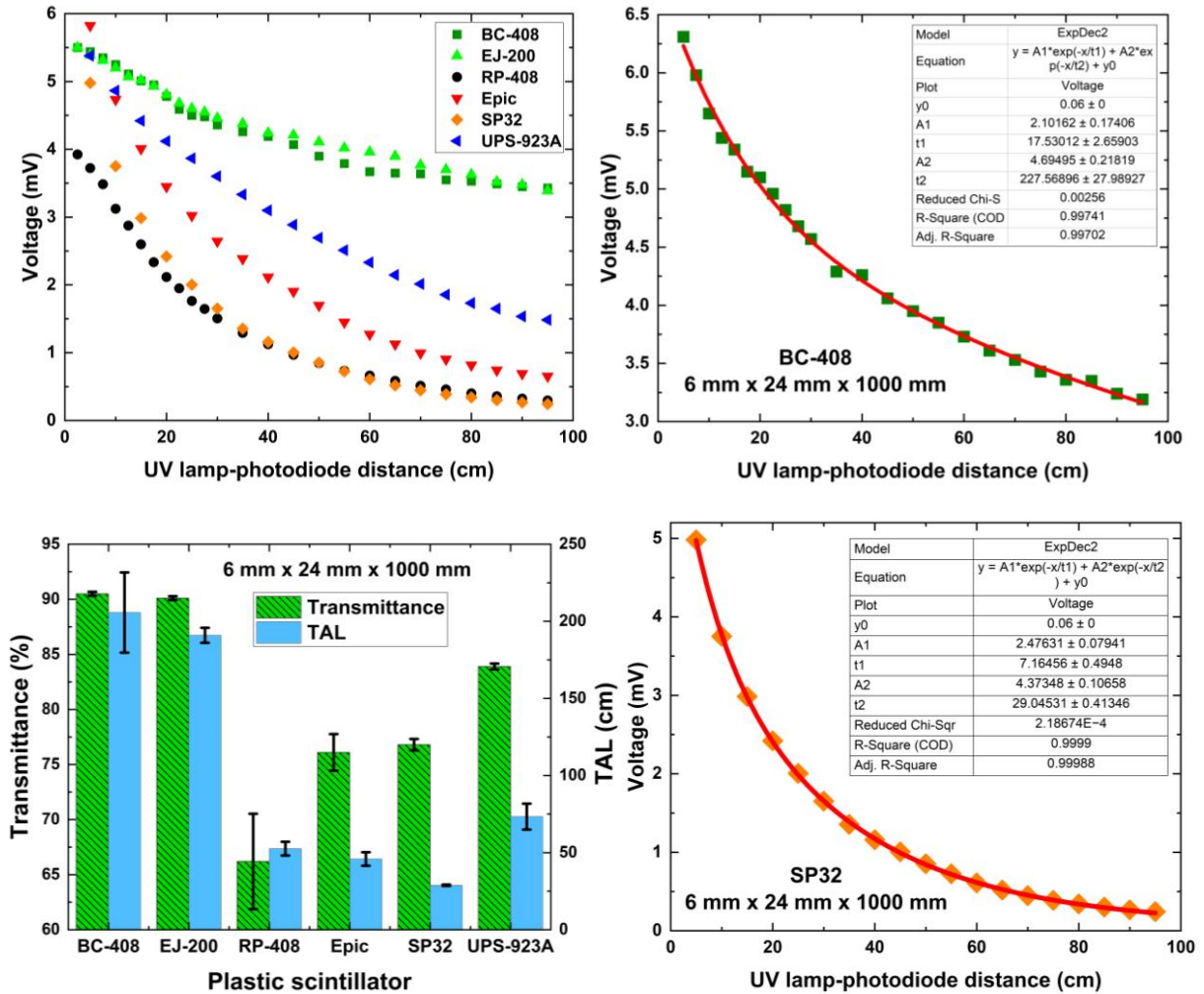

Figure 4. Results from the TAL measurements of commercial plastic scintillators with dimensions 6 mm x 24 mm x 1000 mm. Comparison of the light signal from the measured plastic scintillators converted by photodiode into voltage signals (top left panel), and an example of the fit of sum of two exponential functions (red line) to the measured voltage signals (squares) for BC-408 scintillator (top right panel), and for SP32 scintillator (bottom right panel). In the right panels, parameter y0 denotes background noise from the photodiode and multimeter without UV lamp exciting the scintillator (measured and fixed as average noise to the level of 0.06 mV). Parameter t2 denotes technical attenuation length value in cm unit. Comparison of the transmittance and the TAL values (bottom left panel). Error bars represent standard deviation from five transmittance measurements and four TAL measurements of each scintillator type, respectively.

The results of TAL and transmittance measurements of the plastic scintillators are in good agreement, see bottom left panel in Figure 4. The most transparent plastic scintillators with highest TAL value (the lowest light attenuation) are BC-408 and EJ-200 reaching about 200 cm. The BC-408 and EJ-200 plastic scintillators manufactured with large surfaces as-cast and edges diamond-milled showed the best results in the transmittance and the TAL measurements. Medium result of TAL equals to about 73 cm and it was measured for the UPS-923A scintillator which is also characterized by medium result in the transmittance measurements. The lowest TAL values for the other scintillators are in the range from 29 cm to 52 cm, see Table 3. The TAL values are comparable to the values measured in other experiments, as showed in Table 1. Large standard deviation for BC-408 TAL can be caused by slightly different properties of two bars: scratches on surface, thickness tolerance along 1000 mm long bar, and reflection properties of diamond-milled edges with semicircular pattern. High standard deviation for RP-408 transmittance can be derived from very uneven polishing quality of the surfaces.

The low TAL value reveals high attenuation of the emitted blue light in the scintillator strip. Length of plastic scintillators in radiation detection modules in TOF counters and J-PET scanners should not exceed the TAL value. If plastic scintillators will have length higher than the TAL value, the blue light signal emitted by impinging particles will have low intensity, and resolution of the radiation detector will worsen [15]. For BC-408 scintillator with four different cross-sections, the higher technical attenuation length of the scintillator, the better timing resolution of radiation detector [31]. High TAL values allow to construct long modules of the total-body J-PET scanner or achieve better timing resolution in shorter modules because more light reaching light detectors glued at both ends of the scintillator will increase performance of the J-PET scanner.

### 3.4. Cross-sections of the scintillator

From the above measured six types of plastic scintillators, the EJ-200 with the highest transparency to the visible light and the highest TAL value, was chosen to further research. Additional 1000 mm long strips with cross-section of 6 mm x 12 mm and 6 mm x 6 mm were studied to compare their TAL values to the previously measured EJ-200 scintillator with 6 mm x 24 mm cross-section. Motivation for investigating smaller cross-sections is to measure influence of scintillator dimensions on optical properties of the scintillators which will change timing and spatial resolution of the total-body J-PET scanner. The influence of cross-section of EJ-200 plastic scintillator on the light signal intensity and the TAL value is presented in Table 4 and Figure 5.

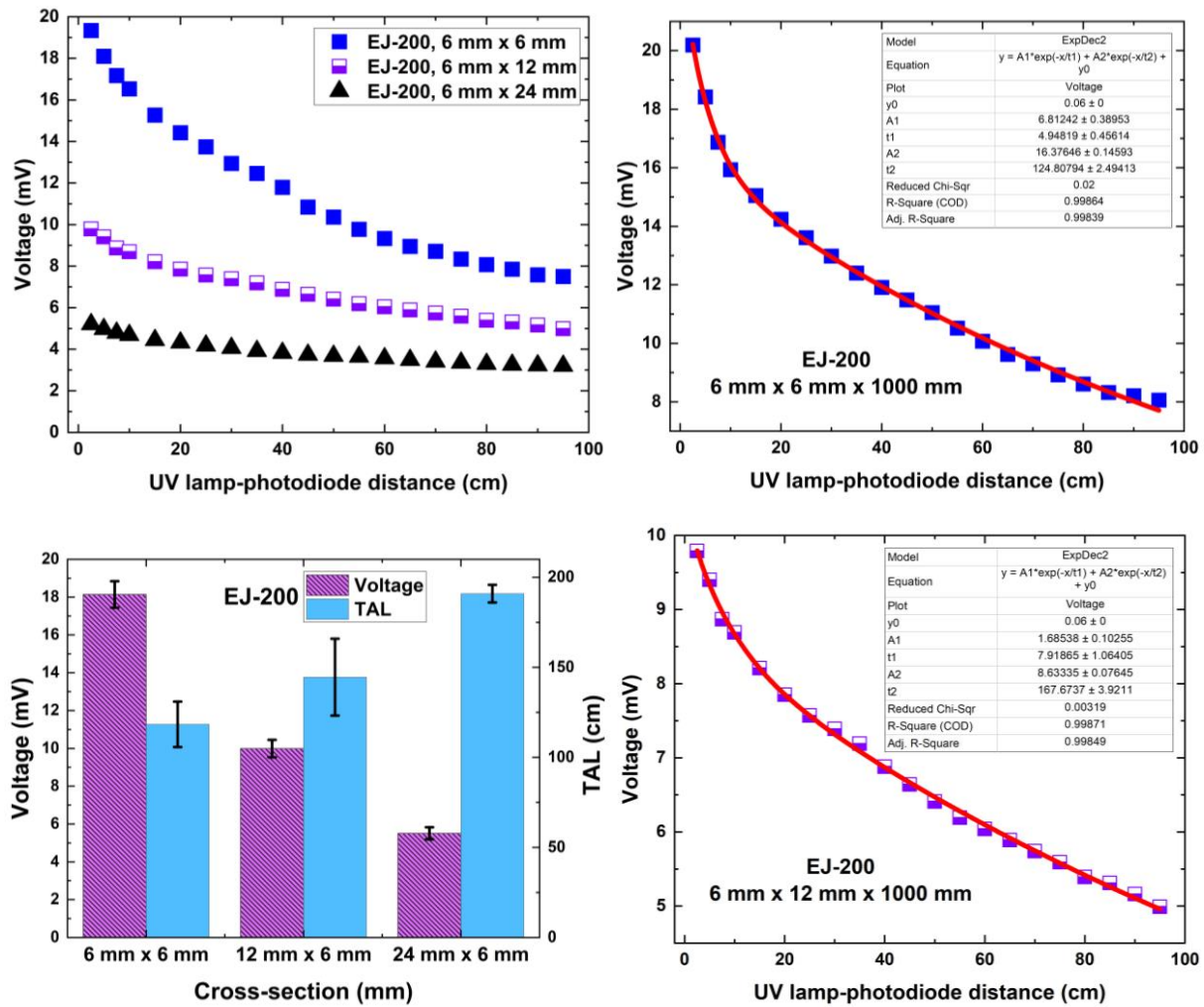

Figure 5. Results from the TAL and voltage signal measurements of commercial plastic scintillator EJ-200 with 1000 mm length and three rectangular cross-sections of 6 mm x 24 mm, 6 mm x 12 mm, and 6 mm x 6 mm. Comparison of the light signal from the measured plastic scintillators converted by photodiode into voltage signals (top left panel), an example of the fit of sum of two exponential functions (red line) to the measured voltage signals (blue squares) for EJ-200 scintillator with 6 mm x 6 mm cross-section (top right panel) and 12 mm x 6 mm cross-section (bottom right panel). Parameter t2 denotes technical attenuation length value in cm unit. Results of TAL and voltage signals for investigated cross-sections (bottom left panel). Error bars for TAL and voltage represent standard deviation from the ten measurements of each cross-section.

During the TAL measurements of EJ-200 plastic scintillators with different cross-sections, photodiode reading the blue light signal was positioned in the center of the 6 mm x 6 mm, 6 mm x 12 mm and 6 mm x 24 mm strip edges at the one end of the scintillator. These positions are similar to the SiPM design in the total-body J-PET scanner, where up to four SiPMs with 6 mm x 6 mm active area can be glued to one end of the plastic scintillator strip. The results of the TAL measurements is given in Table 4 and Figure 5.

Table 4. Results of TAL measurements for 1000 mm long EJ-200 plastic scintillators with three rectangular cross-sections.

| Cross-section (mm x mm) | Average voltage signal 5 cm from photodiode (mV) | Technical attenuation length (cm) |
|---|---|---|
| 6 x 6 | 18.14 ± 0.70 | 118.3 ± 12.6 |
| 12 x 6 | 10.00 ± 0.46 | 144.5 ± 21.4 |
| 24 x 6 | 5.51 ± 0.32 | 190.9 ± 4.9 |

The voltage signal amplitude measured with photodiode has the higher value for 6 mm x 6 mm cross-section. General trend can be observed: decreasing cross-sections of the plastic scintillator increases signal amplitude. This is caused by active to total cross-section ratio covered by photodiode. The scintillators were excited by 5 mm diameter spot of UV light positioned at the center of large surfaces (6, 12 or 24 mm x 1000 mm) of the scintillator. UV light intensity was constant, and amount of the blue light emitted by the scintillator was also constant due to the same 6 mm thickness of the scintillator. The emitted blue light was guided in the scintillator by total internal reflections to the end of the strip where photodiode with radiant sensitive area of 2.73 mm x 2.73 mm was air-coupled at the center of the edge.

The main reason for differences in the signal intensity of scintillators with different cross-sections was ratio of the amount of blue light per surface of the cross-sections. The amount of the blue light reaching the end of the scintillator strip is higher for smaller cross-sections. Higher light intensity for smaller cross-section is connected with the light concentration on the active part of the photodiode – more light photons per square centimeter is guided by the scintillator, and signal is stronger. For the smallest 6 mm x 6 mm cross-section, the same amount of blue light emitted in the UV irradiation spot was concentrated on smaller surface at the end of the strip, and in turn generated stronger voltage signal in the photodiode. For the largest 6 mm x 24 mm cross-section, the same amount of blue light emitted in the UV spot was spread out into larger surface at the end of the strip, and in turn generated weaker voltage signal in the photodiode.

Photodiode with 2.73 mm × 2.73 mm optical sensitive area dimensions smaller than 6 mm thickness of measured scintillators was used to avoid errors from misalignment during air coupling: photodiode was positioned always in the center of scintillator cross-section. In the article [15] we have shown that the photons reaching the end of the scintillator strips are homogenously distributed. Light guide collecting light from scintillators cross-section to active area of photodiode was avoided. We would like test scintillators with similar conditions as in J-PET scanner: SiPM photoactive shape will be 6 mm × 6 mm square without light guide. The light guide decreases light signal due to introducing into setup two additional surfaces manufactured from material with different refractive index than scintillators refractive index.

Obtained TAL values equal to about 190 cm, are the highest for the largest 6 mm x 24 mm cross-section. The TAL value of about 118 cm and 144 cm are measured for the 6 mm x 6 mm and 6 mm x 12 mm cross-sections, respectively. For the smaller cross-sections of the EJ-200 plastic scintillator, attenuation of the emitted light is higher. The higher light attenuation for smaller cross-sections is connected with increasing number of light reflections from the scintillator surfaces, and the loss of light on imperfections on the scintillator surface, as was also measured in [31]. All three cross-sections of EJ-200 scintillator with TAL values over 100 cm are sufficient for construction of J-PET modules. To obtain smaller variation in signal intensity and better time resolution, plastic scintillators with length of 330 mm will be used in next scanner prototypes. To choose optimal cross-section, measurement of resolution with SiPMs attached to scintillators will be performed.

## 4. Conclusions

The maximum emission wavelengths of the measured plastic scintillators range from 420 nm to 429 nm in the blue part of the visible light spectrum. Transparency to visible light of six types of commercial plastic scintillators was measured by two methods: transmission spectra through 6 mm thick scintillator and technical attenuation length of 1000 mm long scintillators. The plastic scintillators with the highest transmittance and the lowest attenuation of emitted blue light are BC-408 and EJ-200. Transmittance at the wavelength of maximum emission for 6 mm thick best scintillators is about 90% and is significantly higher than for four other types of investigated scintillators with the transparency ranging from 66% to 84%. The highest light attenuation length of 206 and 191 cm was found for BC-408 and EJ-200 scintillators, respectively. Measured TAL for two the most transparent scintillators is substantially higher than for other researched scintillators with TAL values in range from 29 cm to 73 cm. From the most transparent plastic scintillators, up to 2 meters long modules of total-body J-PET scanners can be build. First results from the measurements of plastic scintillators with different cross-sections suggest that small cross-section decrease TAL. Decreasing cross-section of EJ-200 scintillator from 24 mm x 6 mm to 6 mm x 6 mm decreases TAL values from 191 cm to 118 cm.

## Acknowledgements

The authors acknowledge support by the TEAM POIR.04.04.00-00-4204/17 program, Ministry of Education and Science through grant SPUB/SP/490528/2021, the National Science Centre grants no. 2019/35/B/ST2/03562, 2021/42/A/ST2/00423, 2021/43/B/ST2/02150, the Jagiellonian University via project CRP/0641.221.2020, and the SciMat and qLife Priority Research Areas budget under the program Excellence Initiative - Research University at the Jagiellonian University.